\documentclass[twocolumn, secnumarabic,amssymb, aps]{revtex4-1}

\usepackage[columnwise]{lineno}

\usepackage[utf8]{inputenc}
\usepackage{CJKutf8}
\usepackage{amsmath,amssymb}
\usepackage{type1cm}
\usepackage{mathrsfs}
\usepackage{mathtools}
\usepackage{url}
\usepackage{graphicx}
\usepackage{color}
\usepackage{booktabs}
\usepackage{braket}

\newcommand\HLPOFR[1]{{\color{black}#1}}

\newcommand\HLPOFRR[1]{{\color{black}#1}}
\usepackage{bm}
\newcommand{\vect}[1]{\boldsymbol{\mathbf{#1}}}
\def\vec#1{\vect{#1}}

\begin{document}
\title{Numerical model of the Gross--Pitaevskii equation for rotating Bose--Einstein condensates using smoothed-particle hydrodynamics}

\author{Satori Tsuzuki~(\begin{CJK}{UTF8}{min}都築怜理\end{CJK})}
\email[Email:~]{tsuzukisatori@g.ecc.u-tokyo.ac.jp\\ \url{https://www.satoritsuzuki.org/} }
\affiliation{Research Center for Advanced Science and Technology, University of Tokyo, 4-6-1, Komaba, Meguro-ku, Tokyo 153-8904, Japan}
\begin{abstract}
This study proposed a new numerical scheme for vortex lattice formation in a rotating Bose-Einstein condensate (BEC) using smoothed particle hydrodynamics (SPH) with an explicit real-time integration scheme. Specifically, the Gross--Pitaevskii (GP) equation was described as a complex representation to obtain a pair of time-dependent equations, which were then solved simultaneously following discretization based on SPH particle approximation. We adopt the 4th-order Runge--Kutta method for time evolution. We performed simulations of a rotating Bose gas trapped in a harmonic potential, showing results that qualitatively agreed with previously reported experiments and simulations. The geometric patterns of formed lattices were successfully reproduced for several cases, for example, the \HLPOFRR{hexagonal} lattice observed in the experiments of rotating BECs. Consequently, it was confirmed that the simulation began with the periodic oscillation of the condensate, which attenuated and maintained a stable rotation with slanted elliptical shapes; however, the surface was excited to be unstable and generated ripples, which grew into vortices and then penetrated the inside the condensate, forming a lattice. We confirmed that each branch point of the phase of wavefunctions corresponds to each vortex. These results demonstrate our approach at a certain degree of accuracy. In conclusion, we successfully developed a new SPH scheme for the simulations of vortex lattice formation in rotating BECs.
\end{abstract}
\maketitle

\section{Introduction}
Vortex lattice phenomena observed in rotating Bose-Einstein condensates (BECs) have garnered attention in many fields related to condensed matter physics. Clarifying the essential mechanism of lattice formation in BECs can significantly contribute to these disciplines. The numerical analyses of the lattice formation process in rotating superfluid helium-4 or ultracold atomic gases have been major topics in this field. Several computational schemes for the direct simulation of the Gross-Pitaevskii (GP) equation, a nonlinear Schr${\rm \ddot{o}}$dinger equation for interacting bosons, have been developed over the past decades. Specifically, mesh-based approaches, such as, the finite difference (FD)~\cite{PhysRevE.62.1382, PhysRevE.62.2937, PhysRevE.62.7438} or finite element (FE)~\cite{VERGEZ2016144, doi:10.1137/15M1009172, HEID2021110165} methods, have frequently been adopted for the spatial discretization of the GP equations. Employing a high-order time-integrating scheme is essential to ensure numerical stability for the iterative computation over a long period; hence, the implicit time-integration schemes using the Crank-Nicolson~\cite{PhysRevE.62.2937, doi:10.1063/1.4887568, MURUGANANDAM20091888} or alternating direction implicit (ADI)~\cite{WANG20161114, LI201538} have been employed. Because the implicit approaches result in matrix computations for solving eigenvalue problems, they introduce the unit of the imaginary time and algebraically update the state of the system. Alternative methods for solving the eigenvalue problem for the time-dependent Schr${\rm \ddot{o}}$dinger equation include the imaginary-time propagation methods~\cite{LEHTOVAARA2007148, doi:10.1063/1.4821126, GOLDBERG1967433, PhysRevLett.82.4956, PhysRevA.61.011601} and the Fourier spectra method~\cite{Muruganandam_2003}. Till date, these numerical approaches have achieved a certain success in capturing the complex dynamics of lattice formation in BECs. 

\HLPOFR{Because the GP equation is a nonlinear Schr${\rm \ddot{o}}$dinger equation for interacting bosons, its Hamiltonian has the form of a many-particle interacting system~\cite{Rogel-Salazar_2013, Salasnich2017BSIUA, doi:10.1063/5.0122247}. By describing the GP equation in numerical simulations as a many-particle interaction system similar to the picture of BEC physics, one can expect a more accurate reproduction of microscopic interactions among atomic particles using analytical particles in the computational domain. However, this is challenging with Eulerian models, such as FE or FD, because discretized points have no physical meaning as particles. Therefore, Lagrangian models have an advantage over these Eulerian models.}
Unfortunately, to the best of our knowledge, no study has reported on the Lagrangian particle approximation to discretize the GP equation to reproduce quantum lattice. The vortex filament method (VFM) is employed for local interaction calculations among several quantum vortices on the basis of the Biot--Savart law~\cite{Idowu2001, doi:10.1063/1.4828892, PhysRevLett.120.155301, doi:10.1063/1.5091567, PhysRevLett.124.155301}; however, currently, it is not suitable for the collective dynamics simulations of the condensates because of the difference in physical scales. In contrast, smoothed particle hydrodynamics (SPH)~\cite{gingold1977smoothed}, which originated in astrophysics, is used for fluid dynamics calculations and for general purposes such as a finite particle approximation for the partial differential equation~\cite{Liu2010, imoto2019convergence, doi:10.1063/1.5068697}. SPH can be expected to work as a Lagrangian particle numerical scheme for the direct simulations of the GP equation.
In addition, a recent study~\cite{doi:10.1063/5.0060605} suggested that a two-fluid model for superfluid helium-4 can be solved using SPH to produce vortex lattices under specific conditions. Another study~\cite{doi:10.1063/5.0122247} reported a theoretical proposal that the motion equation for inviscid fluid in the two-fluid model becomes equal to the quantum fluid equation derived from the GP equation under specific conditions, in case of SPH formalism. Therefore, simulating the behavior of the GP equation in SPH form may contribute to finding certain connection to these studies.

This study proposed a new numerical scheme for vortex lattice formation in a rotating BEC using SPH with an explicit real-time integration scheme. Specifically, the Gross--Pitaevskii (GP) equation was proposed in complex representation to obtain a pair of time-dependent equations, which were obtained in the real and imaginary parts, respectively. These equations were then discretized and solved based on SPH particle approximation. We adopted the 4th-order Runge--Kutta method for time evolution; our scheme has the accuracy of the second order in space and the fourth order in time evolution. 
The results of SPH simulations of a rotating Bose gas trapped in a harmonic potential were qualitatively consistent with previously reported experiments and simulations. Notably, we succeeded in reproducing the geometric patterns of formed lattices for several cases, for example, the \HLPOFRR{hexagonal} lattice observed in the experiments of rotating BECs. 
We also report that the simulation began with the periodic oscillation of the condensate, which attenuated and maintained a stable rotation with slanted elliptical shapes; however, the surface was excited to be unstable and generated ripples, which grew into vortices and then penetrated inside the condensate, forming a lattice. We confirmed that each branch point of the phase of wavefunctions corresponded to each vortex. Meanwhile, the Lagrangian approaches such as SPH incur huge computational costs. Thus, we developed an explicit real-time integration scheme for SPH to be compatible with computational accelerators such as graphics processing units (GPU). Consequently, this study demonstrated that the Lagrangian numerical models with explicit time integration can reproduce the dynamics of vortex-lattice formation by selecting the appropriate scheme. 

The remainder of this paper is structured as follows. In Section~\ref{sec:methods}, we explain the targeted governing equation and its discretized form in SPH. We present the several numerical tests for vortex lattice formation in a rotating BECs in Section 3 and discusses the simulation results in Section 4. Finally, Section 5 concludes this paper.

\section{Methods} \label{sec:methods}
\subsection{Gross--Pitaevskii equation in the complex representation} \label{seq:GPtheory}
The dynamics of a rapidly rotating Bose gas trapped in a harmonic potential can be described by the following two-dimensional GP equation on condition that the z (vertical) component of the condensed wavefunction can be separated from the other components as follows~\cite{PhysRevA.67.033610}:
\begin{eqnarray}
(i-\gamma)\frac{\partial \psi}{\partial t} = \Biggl[ -\nabla^2 + \frac{1}{4}\{(1+\epsilon_{x}) x^{2}+(1+\epsilon_{y}) y^{2}\} \nonumber \\
+ C|\psi|^2 -\mu + i\Omega({x}\partial_{y}-{y}\partial_{x})\Biggr]\psi, \label{eq:2dGPeq}
\end{eqnarray}
where $\epsilon_x$ and $\epsilon_y$ indicate the anisotropy parameters of the harmonic potential, $\mu$ is chemical potential, and $\gamma$ is the phenomenological dissipation~\cite{PhysRevA.65.023603}. 
Let us denote the angular frequency in the x or y (horizontal) direction as $\omega_{\perp}$ and that in the z (vertical) direction as $\omega_{z}$. When the ratio of $\omega_{\perp}$ to $\omega_z$, $\lambda$, is sufficiently larger than one ($\lambda =\omega_{\perp}/\omega_{z} \gg 1$), the parameter $C$ is expressed as $C= 4\sqrt{\pi\lambda}N a/a_h$, where $N$ is the number of bosonic particles projected to the two-dimensional plate, $a$ is the scattering length, and $a_{h}$ represents the unit of length.
Equation~(\ref{eq:2dGPeq}) is the dimensionless version of the original GP equation; the length, time, and wavefunction are scaled as $\bar{x} = a_{h}x$, $\bar{t} = \omega_{\perp}^{-1} t$, and $\bar{\psi} = \sqrt{N}a_{h}^{-1} \psi$, respectively. Here, $\bar{x}$, $\bar{t}$, and $\bar{\psi}$ are the variables of the original GP equation. Moreover, $\Omega$ in Eq.~(\ref{eq:2dGPeq}) indicates the scalar multiple of $\omega_{\perp}$. 

Let us describe Eq.~(\ref{eq:2dGPeq}) in the complex representation as a preparation for the SPH calculation with explicit time-integrating scheme. By substituting the complex representation of wavefunction $\psi$, $\psi=u+iw$, into Eq.~(\ref{eq:2dGPeq}) and comparing each part of the complex expression on the left and right sides, the following simultaneous equations are obtained after simple calculations:
\begin{eqnarray}
\frac{\partial X}{\partial t} &=& \Omega(x\partial_{y}{u} - y\partial_{x} u) \nonumber \\
		&~& - [\nabla^2{w} - (C|\psi|^2 + V - \mu){w}], \nonumber \\
\frac{\partial Y}{\partial t} &=& \Omega(x\partial_{y}{w} - y\partial_{x} w) \nonumber \\
	    &~& + [\nabla^2{u} - (C|\psi|^2 + V - \mu){u}], \nonumber \\ 
X~&\coloneqq&~ u - \gamma{w}, \nonumber \\
Y~&\coloneqq&~ \gamma{u} + w, \nonumber \\
V~&\coloneqq&~ \frac{1}{4}\{(1+\epsilon_{x}) x^{2}+(1+\epsilon_{y}) y^{2}\}. \label{eq:simuleq} 
\end{eqnarray}
Here, $|\psi|^2=u^2+w^2$, where $\psi$ must satisfy the following normalization conditions all the times:
\begin{eqnarray}
\int |\psi|^2 dxdy = 1. \label{eq:normalcond} 
\end{eqnarray}
\HLPOFR{
Equation~(\ref{eq:normalcond}) represents the law of quantum mechanics, which states that the integral of the square of wavefunction over the entire spatial domain is always one~\cite{Berman2018}.} 
To perform the simulations, we adopted the 4th-order Runge-Kutta method for the time-integration of variables $X$ and $Y$ in Eq.~(\ref{eq:simuleq}). In addition, we normalized the wave functions according to Eq.~(\ref{eq:normalcond}) at every time step of the simulation. The SPH representation of the operators on the right-hand side of Eq.~(\ref{eq:simuleq}) is explained in Section~\ref{sec:explainsph}. 

\subsection{An overview of smoothed-particle hydrodynamics} \label{sec:explainsph}
The SPH was first described in astrophysics~\cite{gingold1977smoothed, monaghan1992smoothed} and is now widely used as a finite particle approximation method for continuous distributions. It exploits the fact that a discrete physical quantity $\varphi$ can be expressed as a continuous quantity using the Dirac delta function $\delta$ ($\varphi(\vec{r}) = \int \varphi(\vec{r})\delta(\vec{r} - \vec{\acute{r}}) d\vec{\acute{r}}$). In SPH, the $\delta$ function is approximated using a distribution function $W$, referred to as the kernel function as 
\begin{eqnarray}
\varphi(\vec{r}) \simeq \int \varphi(\vec{r})W(\vec{r} - \vec{\acute{r}}, h) d\vec{\acute{r}}. \label{eq:sphdef}
\end{eqnarray}
Here, $W$ exhibits the properties of smoothness, symmetry around an axis ($W(\vec{r}) = W(-\vec{r})$), and normalization ($\int W d\vec{r} = 1$). Moreover, it converges to the $\delta$ function when the kernel radius $h$, or distribution width, approaches zero ($\lim_{h \to 0} W(\vec{r} - \vec{\acute{r}}, h) = \delta (\vec{r} - \vec{\acute{r}})$). A straightforward example of $W$ satisfying the aforementioned conditions is the Gaussian kernel~\cite{gingold1977smoothed} represented as $W(\vec{r} - \vec{\acute{r}}) = C_{\rm sph}/h^d {\rm exp}[-|\vec{r} - \vec{\acute{r}}|^2 /h^2]$, where $C_{\rm sph}$ denotes a normalization constant, $h$ denotes the kernel radius, and $d$ represents the dimension. 

Equation~(\ref{eq:sphdef}) can be written in the discretized form considering summation approximation as 
\begin{eqnarray}
\varphi(\vec{r}_i) = \sum_{j=1}^{N_p} \varphi(\vec{r}_j) {\Delta{V}_j}W_{ij}, \label{eq:sphdisc}
\end{eqnarray}
where $W_{ij} = W(|\vec{r}_{i}-\vec{r}_{j}|, h)$ and $\Delta{V}_j$ represents the discretized small volume.
After operating $\nabla$ to Eq.~(\ref{eq:sphdisc}) from the left, a simple calculation using vector analysis and Gauss' divergence theorem yields the gradient and Laplacian of $\varphi$ as follows~\cite{monaghan1992smoothed}:
\begin{eqnarray}
\nabla \phi(\vec{r}_{i}) 
	&=& \sum^{N_{p}}_{j} \Delta{V}_{j} 
	\Bigl[ \phi(\vec{r}_{i}) + \phi(\vec{r}_{j}) \Bigr] 
			\nabla W_{ij},\label{eq:gradient} \\
\nabla^{2} \phi(\vec{r}_{i}) 
	&=& \sum^{N_{p}}_{j} \Delta{V}_{j} \frac{\phi(\vec{r}_{i})-\phi(\vec{r}_{j})}{|\vec{r}_{i}-\vec{r}_{j}|^{2}} 
			(\vec{r}_{i} - \vec{r}_{j})\cdot \nabla W_{ij},\label{eq:laplacian}
\end{eqnarray}
where $\nabla W_{ij}$ denote the gradient of $W_{ij}$. Equations (\ref{eq:gradient}) and (\ref{eq:laplacian}) were derived assuming that each fragment has the same uniform volume from the standard forms in~\cite{monaghan1992smoothed}, wherein the volumes of the $i$th and $j$th fragments are expressed separately. This simplification is valid for this study because all the fragments, which are the discrete points of volume, are fixed at the same locations as in the initial state during simulations \HLPOFR{owing to that the governing equation in Eq.~(\ref{eq:2dGPeq}) has no advection term. In other words,} this study presents a static model that describes the system in SPH form \HLPOFR{and simultaneously allows us} to ensure numerical accuracy.

\HLPOFR{
In the first equation of Eq.~(\ref{eq:simuleq}), we name the first and second terms as the rotational and oscillatory terms, respectively. We refer to $(C{|\psi|}^{2} + V - \mu)$ in the parentheses of the oscillatory term as the potential term. The same terminology is applied to the second equation of Eq.~(\ref{eq:simuleq}). In the repeated process of the first equation of Eq.~(\ref{eq:simuleq}), we first update the potential term using the positions and the current values of $u$ and $w$. Next, we calculate the oscillation term by summing the product of the potential term with $w$ and the Laplacian of $w$. Here, the Laplacian of $w$ has been obtained by substituting $w$ for $\phi$ in Eq.~(\ref{eq:laplacian}). We also calculate the rotational term by using the position and gradient of $u$, which was obtained by substituting $u$ for $\phi$ in Eq.~(\ref{eq:gradient}), where $\partial_{a} \phi$ represents the $a$-component of the gradient $\nabla {\phi}$. We update the variable $X$ after obtaining the right-hand side, the difference between the rotational and oscillatory terms. We update the variable $Y$ through a similar process for the second equation of Eq.~(\ref{eq:simuleq}). We then obtain the updated $u$ and $w$ from $X$ and $Y$ using the third and fourth equations of Eq.~(\ref{eq:simuleq}). We iteratively repeat this process in an explicit time-integrating scheme using the 4th-order Runge--Kutta method. Then, the values of $u$ and $w$ are renormalized to satisfy Eq.~(\ref{eq:normalcond}).}

To ensure the numerical accuracy of the Laplacian calculation, Eq.~(\ref{eq:laplacian}) can be rewritten to the form of the Laplacian operator of the moving particle semi-implicit (MPS). This is a particle method where the gradient at a point is estimated by the weighted arithmetic mean of the local gradients defined between the point and other neighboring points in space~\cite{koshizuka1996moving, Koshizuka1998}. In brief, MPS is one form of the finite difference method (FDM) generalized in the Lagrangian form~\cite{imoto2019convergence}. For the detailed explanation of the reformulation of the Laplacian operator, refer to the appendix in Ref.~\cite{Tsuzuki_2021}. In simulations, we applied Eqs (\ref{eq:gradient}) and (\ref{eq:laplacian}) to the first term and the first part in the square bracket of the second term in Eq.~(\ref{eq:simuleq}) after the reformulation. In contrast to the SPH scheme for fluid dynamics, other stabilization techniques or improved SPH operators are not required to compute Eq.~(\ref{eq:simuleq}). 

\section{Numerical analysis}
Numerical simulations of the vortex lattice formations were performed using our SPH scheme. The parameters $(\gamma, C, N, \lambda, a, a_{h}, \omega_{\perp})$ introduced in Section~\ref{seq:GPtheory} were set as $(0.03, 500, 3\times{10}^{5}, 9.2, 5.77, 0.728, 2\pi \times 108.56)$, respectively, to replicate the previous numerical tests based on the FDM-based scheme reported in Ref.~\cite{PhysRevA.67.033610}. Here, the units of the last three parameters $a$, $a_{h}$, and $\omega_{\perp}$ are $\rm nm$, $\rm \mu m$, and $\rm Hz$, respectively. We estimated the chemical potential $\mu$ as $\mu\approx(15/8 N\lambda a/a_{h})^{2/5}$ based on the Thomas--Fermi (TF) model. For the anisotropy parameters $\epsilon_{x}$ and $\epsilon_{y}$, we set $\epsilon_{y}$ to be zero during simulations while rapidly increasing $\epsilon_{x}$ from zero to $\rm 0.05128205128$ in $\rm 20$ msec at the beginning of the simulation; this condition corresponds to the case that the entire anisotropy $\epsilon\coloneqq {(1+\epsilon_{x})-(1+\epsilon_{y})}/{(1+\epsilon_{x})+(1+\epsilon_{y})}$ changes from zero to $\rm 0.025$ as per the numerical tests in Ref.~\cite{PhysRevA.67.033610} and experiments by Madison et al~\cite{PhysRevLett.86.4443}. 
Simulations were performed for different values of $\Omega$: 0.57, 0.7, 0.86, and 0.9 to compare the resulting lattices.

For computational conditions, we set the simulation domain $(L_x, L_y)$ to be $(\rm 40, 40)$. An initial profile was assigned to the real component of the wavefunction according to the profile given by TF approximation around the z-axis with a radius of $\rm 8.85581745325$, which was determined from chemical potential $\mu$. The resolutions in the $x$ and $y$ directions, $(N_x, N_y)$ were set as $\rm (380, 380)$. The same SPH parameters optimized in the numerical tests for a wave propagation problem reported in Section~5.2 in Ref.~\cite{Tsuzuki_2021} were used; the kernel radius $h$ was $0.4d$, where $d$ is the distance between the two nearest discrete points arranged along the orthogonal grids of the two-dimensional space in the initial state. We adopted the Gaussian kernel function. Here, the Gaussian kernel does not exhibit the compact support property, where the kernel exhibits a value of zero at a finite distance from its center point. Thus, a cutoff distance with a radius of $3h$ was introduced. 
\HLPOFR{In addition, we imposed the Dirichlet boundary condition $u = w = 0$ on the edge of the simulation domain.}
\HLPOFR{As} a boundary treatment for the rotational and oscillatory forces, we introduced another cutoff distance $d_c$ at around the outer area of the simulation domain; the forces computed at the distance of $d_c$ or more significant from the center point were attenuated using a Gaussian filter to be smoothly connected to the edge of the domain where the value of wavefunction was always zero. Thus, we set the value of $d_c$ to 15 to ensure the sufficient space for the condensates.

Equation~(\ref{eq:2dGPeq}) assumes the existence of one particle in the ground state in the z-direction, and therefore the z-component of the wavefunction can be described by Gaussian distribution~\cite{PhysRevA.67.033610}. Thus, Eq.~(\ref{eq:2dGPeq}) postulates a quasi-2D situation. We can replicate this situation in our SPH model using a 3D kernel function. As the cutoff length of the kernel function was set to $3h$ and $h=0.4d$, the kernel function contained three layers of discrete points in one direction. Thus, we placed the positive and negative layers parallel to the $xy$-plane at $z=0$ such that the kernel function could hold three layers in the $z$-direction. We then projected their contributions onto the $xy$-plane at $z=0$. These layers can be virtually set because the 2D and 3D Gaussian kernels only differ in the scale of $C_{\rm sph}$, the normalization coefficient of the SPH kernel (Section~\ref{sec:explainsph}). The theoretical value of $C_{\rm sph}$ is obtained as $(\pi{h^2})^{-1}$ for 2D and $({\pi^{3/2}}{h^3})^{-1}$ for 3D calculations~\cite{gingold1977smoothed}. However, because actual calculations include discretization and summation errors, the direct use of these analytical values does not guarantee sufficient accuracy of normalization. We calibrated $C_{\rm sph}$ by referring to the correspondence table between the number of vortices and $\Omega$ reported in Ref.~\cite{PhysRevA.67.033610}. Specifically, we assumed the linear relationship between the value of $\Omega$ and the number of vortices between $\Omega = 0.57$ and $\Omega = 0.9$. Subsequently, $C_{\rm sph}$ was estimated by referring to the results for $\Omega = 0.9$. We then used this value in the remaining three cases of $\Omega = 0.57$, $\Omega = 0.7$, and $\Omega = 0.86$. We set the discrete time $dt$ to be $1.0\times 10^{-4}~{\omega_{\perp}}^{-1}$ and computed $1,000~{\omega_{\perp}}^{-1}$ in total; this is approximately equal to $\rm 1466~msec$ on the original time scale. 
\HLPOFR{We implemented our scheme on a single GPU, NVIDIA GeForce RTX2080 Ti, using C/C++ and CUDA~\cite{4541126}. We utilized several computational techniques to speed up the simulations; Neighbor-particle lists are adopted to reduce the computational cost of finding particles within the radius of interactions from $\mathcal{O}(N^{2})$ to $\mathcal{O}(N)$. We combined the linked list technique~\cite{GREST1989269, GOMEZGESTEIRA2012289} with the neighbor-particle list to reduce memory usage. Although we developed our original calculation code, several open-source frameworks should serve as references for implementation~\cite{GOMEZGESTEIRA2012289, ramachandran2013pysph}.}

\begin{figure*}[t]
\vspace{+0.5cm}
 \includegraphics[width=1.0\textwidth, clip, bb= 0 0 4321 1718]{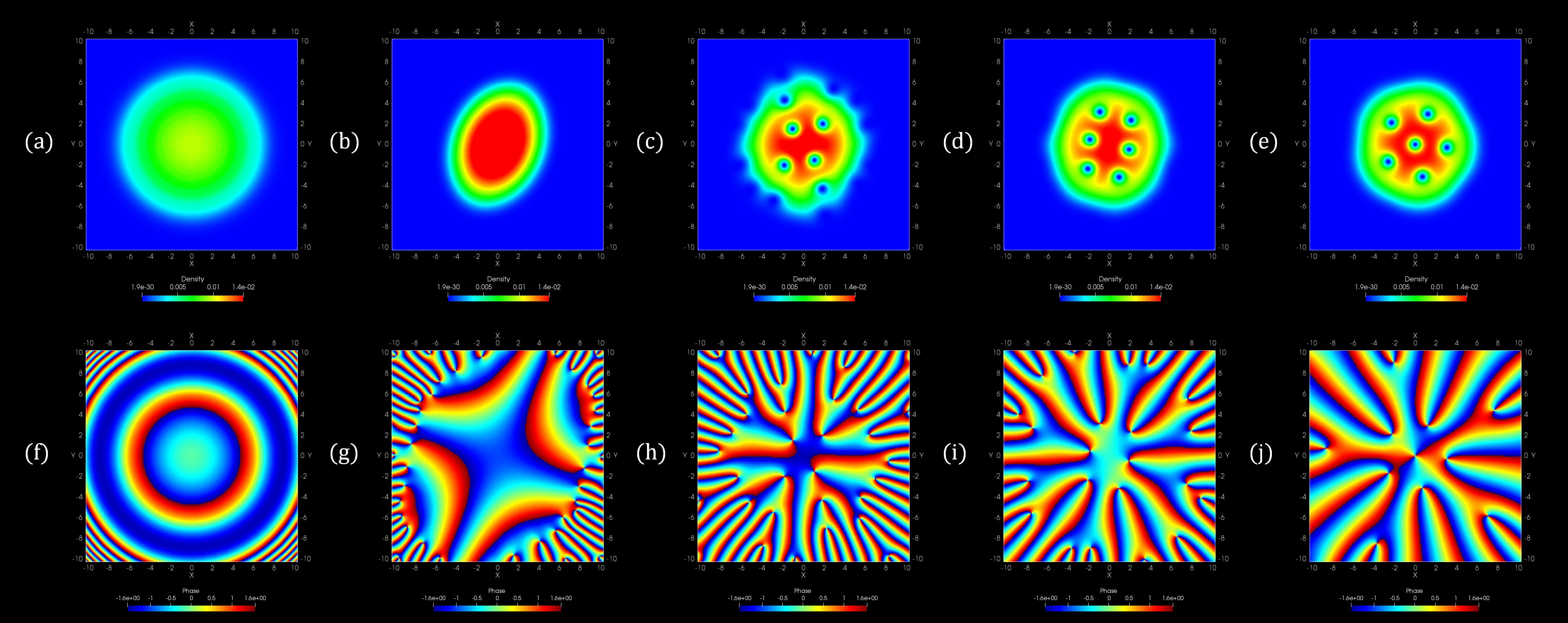}
\caption{The snapshots of the vortex lattice formation obtained by solving the GP equation in Eq.~(\ref{eq:simuleq}) for $\Omega=0.7$. The top row shows the density profiles $|\psi|^2$ approximately at (a) 0.5~msec, (b) 39~msec (c) 244~msec, (d) 391~msec, and (e) 977~msec. The lower rows (f)-(j) show the phase profiles given by $\theta = {\rm atan}(u/w)$ for the corresponding density profiles in (a)-(e), respectively.}
\label{fig:Figure1}
\end{figure*}

\begin{figure*}[t]
\includegraphics[width=1.0\textwidth, clip, bb= 0 0 1686 859]{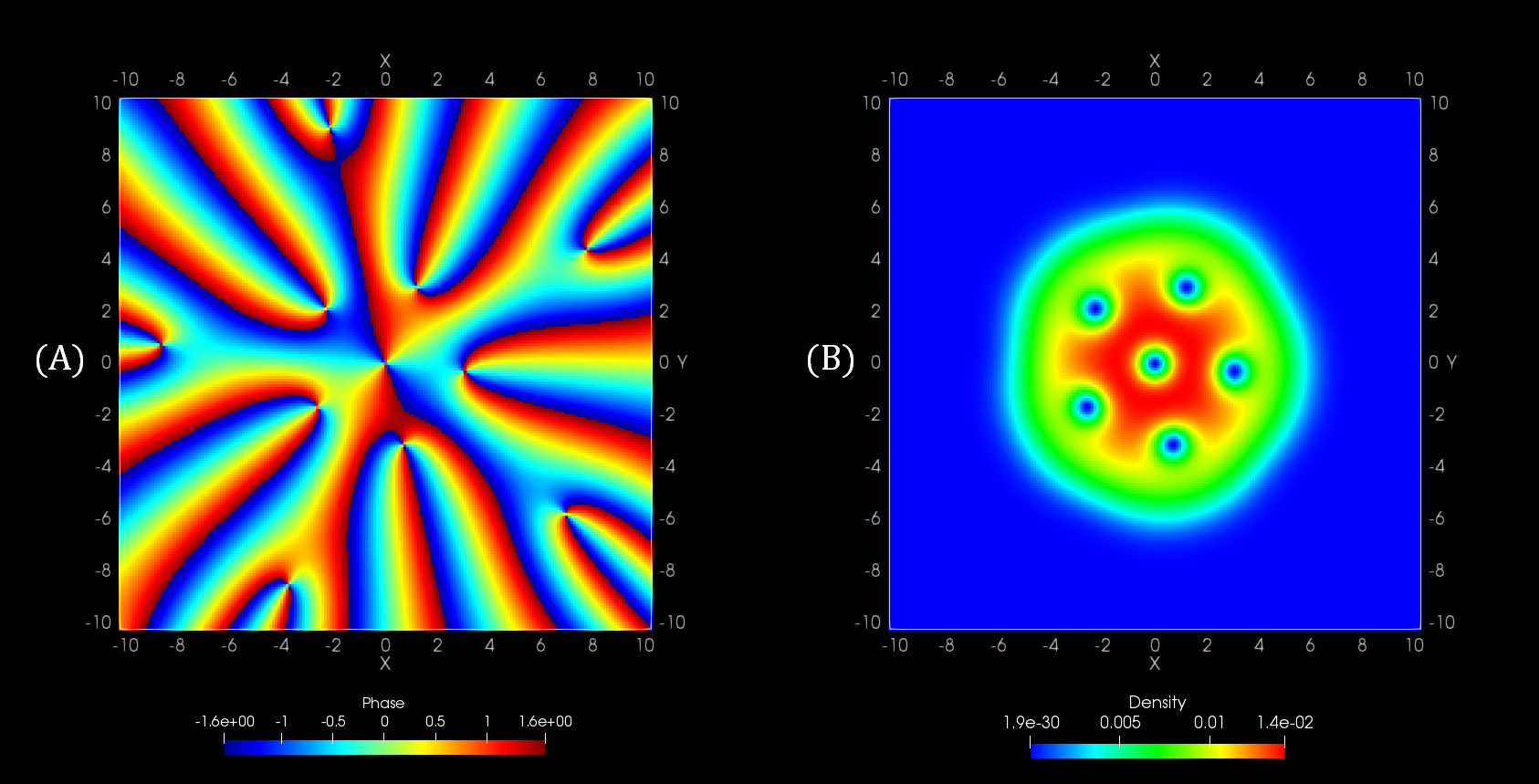}
\caption{A video of (A) the phase and (B) density profiles of the snapshot in Fig.~\ref{fig:Figure1} (Multimedia view) }
\label{fig:Figure2}
\end{figure*}

\begin{figure*}[t]
\vspace{+0.5cm}
\includegraphics[width=0.99\textwidth, clip, bb= 0 0 3915 1786]{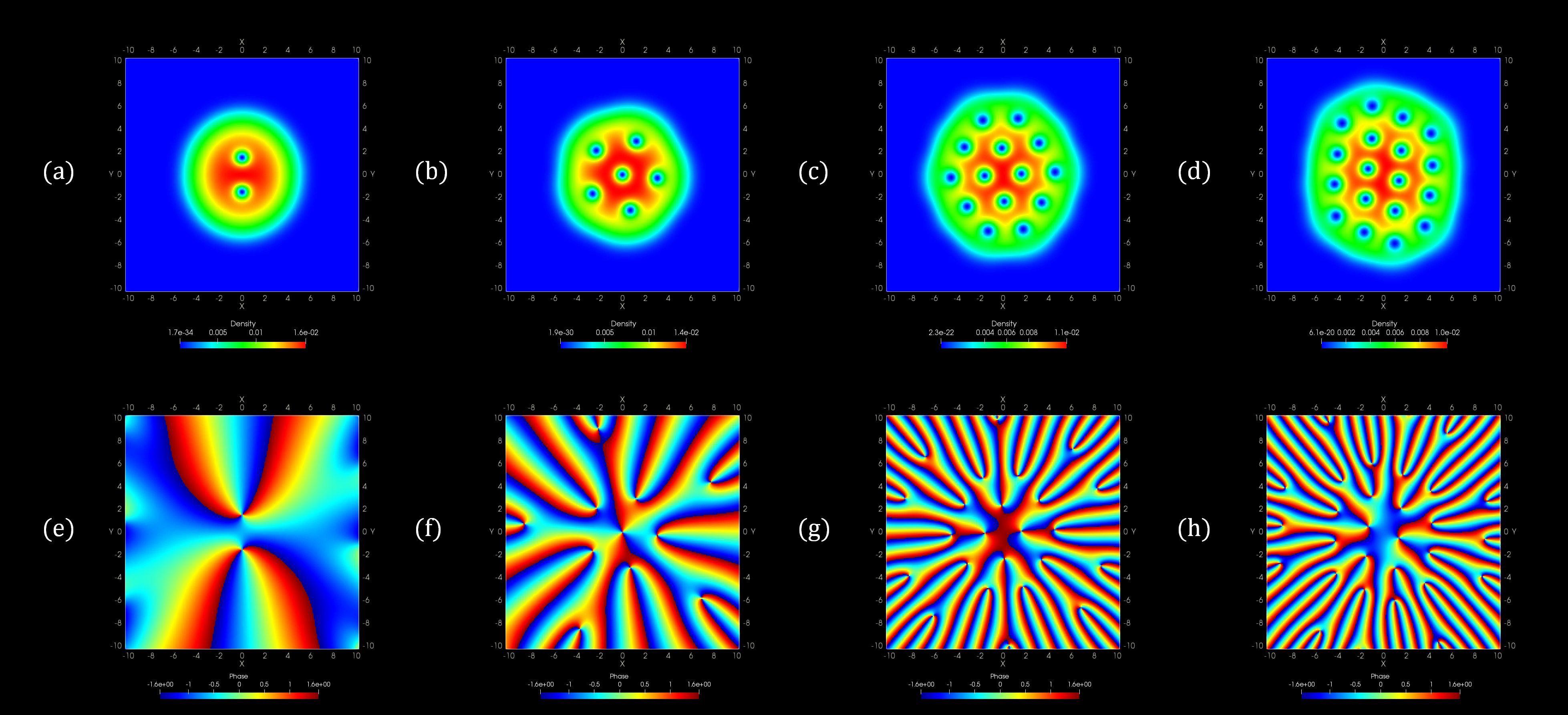}
\caption{The snapshots of vortex lattice formation for the same conditions as in Fig.~\ref{fig:Figure1}, for (a) $\Omega = 0.57$, (b) $\Omega = 0.7$, (c) $\Omega = 0.86$, and (d) $\Omega = 0.9$, for a sufficient time longer than 1400 msec. The lower panels (e)-(h) show the phase profiles for the corresponding density profiles in (a)-(d), respectively.}
\label{fig:Figure3}
\end{figure*}

\begin{figure*}[t]
\includegraphics[width=1.0\textwidth, clip, bb= 0 0 1686 859]{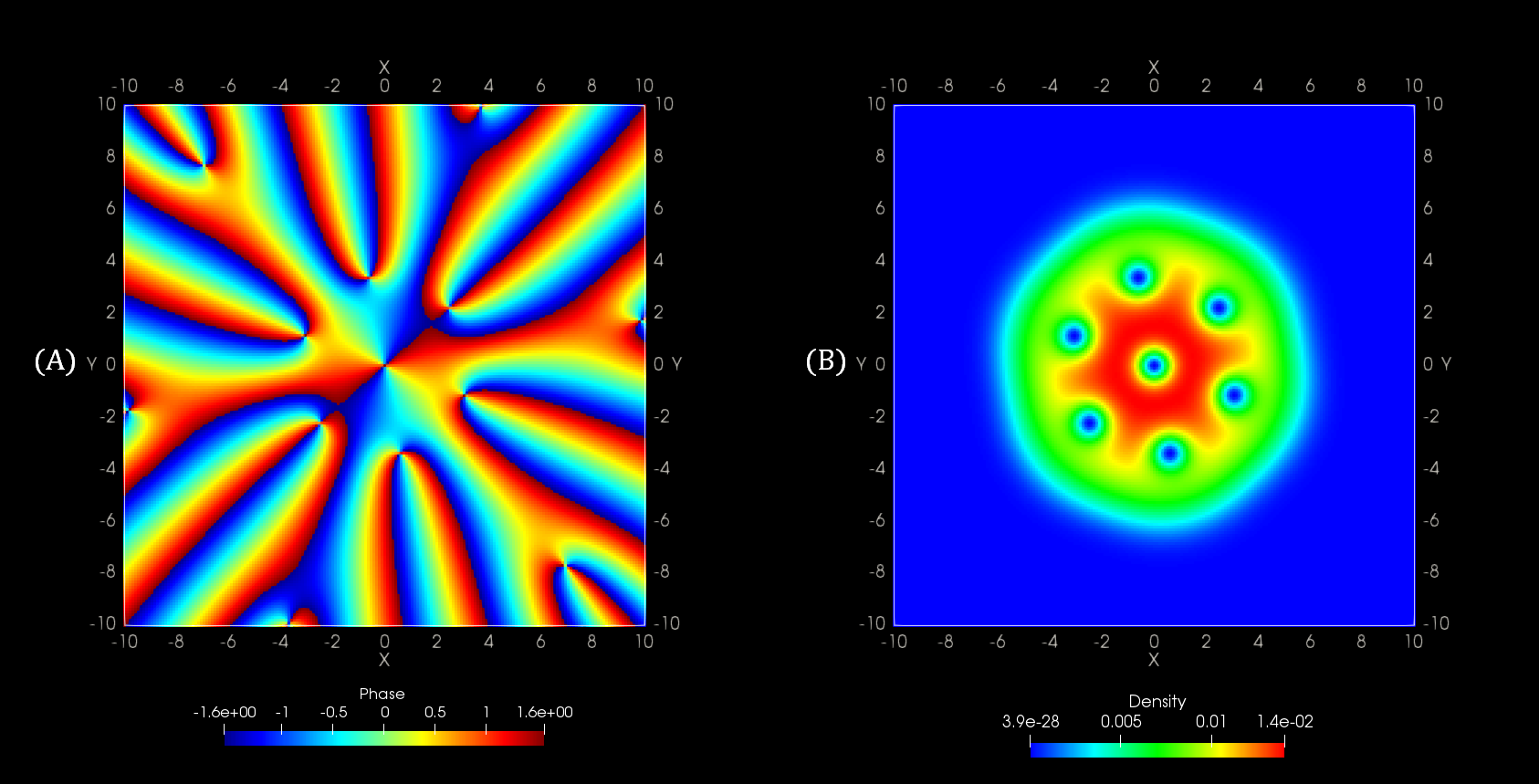}
\caption{The snapshots of (A) the phase and (B) density profiles of a hexagonal lattice for $\Omega = 0.70625$.}
\label{fig:Figure4}
\end{figure*}

Figure~\ref{fig:Figure1} shows the snapshots of an SPH simulation of the vortex lattice formation obtained by solving the GP equation in Eq.~(\ref{eq:simuleq}) for $\Omega=0.7$. The top row in Fig.~\ref{fig:Figure1} displays the density profiles $|\psi|^2$ approximately at (a) 0.5~msec, (b) 39~msec (c) 244~msec, (d) 391~msec, and (e) 977~msec. The lower rows (f)-(j) show the phase profiles obtained as $\theta = {\rm atan}(u/w)$ for the corresponding density profiles in (a)-(e), respectively. 
A video of the snapshots in Fig.~\ref{fig:Figure1} is provided as integral multimedia in Fig.~\ref{fig:Figure2} (ancillary files for the preprint version).
Figure~\ref{fig:Figure3} shows the snapshots of vortex lattice formations under the same conditions as in Fig.~\ref{fig:Figure1}, for different cases of (a) $\Omega = 0.57$, (b) $\Omega = 0.7$, (c) $\Omega = 0.86$, and (d) $\Omega = 0.9$, after a sufficient time longer than 1400 msec. The lower panels (e)-(h) show the phase profiles for the corresponding density profiles in (a)-(d), respectively. 
\HLPOFR{Figure~\ref{fig:Figure4} shows the simulation results for $\Omega = 0.70625$ under the same conditions as in Fig.~\ref{fig:Figure3}, evidently demonstrating that we could reproduce a hexagonal lattice, which is a typical shape for vortex lattices in rotating BECs.}


\begin{figure*}[t]
\vspace{-9.4cm}
\begin{center}
\includegraphics[width=2.0\textwidth, clip, bb= 0 0 1503 807]{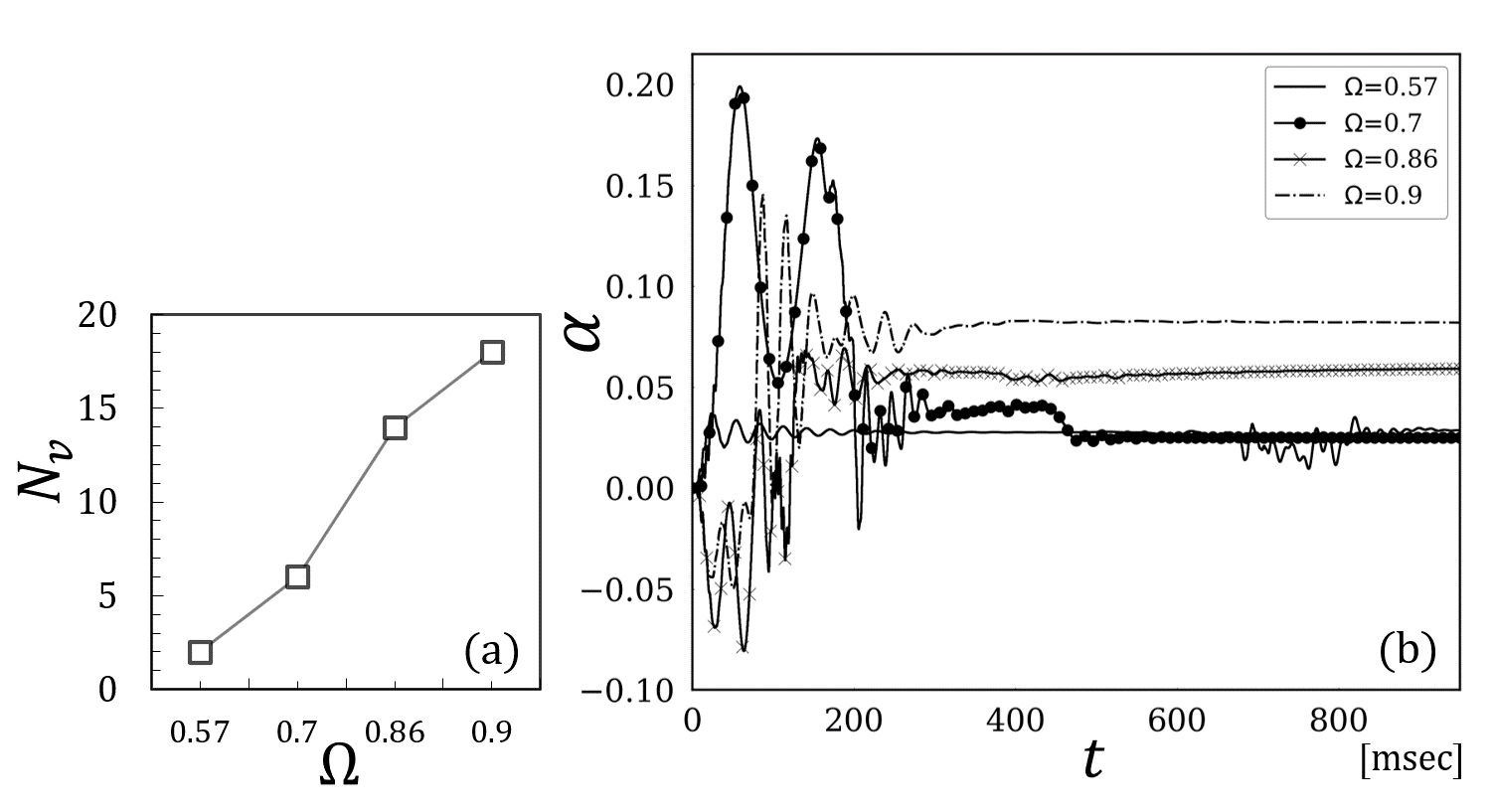}
\end{center}
\caption{Detailed breakdown of the simulations presented in Fig.~\ref{fig:Figure3}: (a) $\Omega$-dependence of the number of vortices, and (b) time-dependence of the deformation parameter $\alpha$ for different cases $\Omega=0.57$, $\Omega=0.7$, $\Omega=0.86$, and $\Omega=0.9$.}
\label{fig:Figure5}
\end{figure*}

The following observations were made in Fig.~\ref{fig:Figure1}. The simulation began with the periodic oscillation of condensate, which gradually attenuated and the condensate maintained a stable rotation with slanted elliptical shapes. However, concurrently, the surface of condensates were gradually excited to be unstable and generated ripples, which grew into vortices and then penetrated into the inside condensates, eventually forming a lattice. 
Notably, we observed a regular \HLPOFRR{single pentagon} after \HLPOFRR{a sufficient time longer than 977 msec}. It was confirmed that the geometric patterns of lattices for all the comparable cases of $\Omega=0.57$, $\Omega=0.7$, and $\Omega=0.86$ in Fig.~\ref{fig:Figure3} were consistent with the numerical tests in Ref.~\cite{PhysRevA.67.033610}, respectively. In addition, the increase in the number of vortices with increase in $\Omega$ was observed, as shown in Fig.~\ref{fig:Figure3}. Figure~\ref{fig:Figure5}(a) shows $\Omega$-dependence of the number of vortices for the simulations displayed in Fig~\ref{fig:Figure3}; the dependence was quantitatively consistent with the numerical results for all the reported cases of $\Omega=0.57$, $\Omega=0.7$, and $\Omega=0.86$ in Ref.~\cite{PhysRevA.67.033610}. 
Furthermore, the phase patterns were confirmed to be well reproduced in all the cases. Here, $u$ and $w$ were the real and imaginary parts of the wavefunction, and hence the phase discontinuities in the lower rows of Figs.~\ref{fig:Figure1} and \ref{fig:Figure3} represent the branch cuts in the complex plane. Through comparisons of the geometrical locations of the quantum vortices in the density profiles presented in the upper rows, it was confirmed that these branch points corresponded to the locations of vortices. 
\HLPOFR{The shape of the hexagonal lattice in Fig.~\ref{fig:Figure4} is confirmed to agree with that of lattices reported in the related work~\cite{PhysRevLett.86.4443, KISHORKUMAR201974}.}
In summary, the key observations specific to the formation of vortex lattices were consistent with numerical~\cite{PhysRevA.67.033610, PhysRevA.65.023603, KISHORKUMAR201974} and experimental results~\cite{PhysRevLett.86.4443}, demonstrating that our approach exhibited a certain degree of accuracy. 

\section{Discussion}
Our simulations were qualitatively consistent with both experiments and simulations in the previous studies~\cite{PhysRevA.67.033610, PhysRevA.65.023603, PhysRevLett.86.4443, KISHORKUMAR201974} in terms of the phenomena characteristic of quantum lattices: the dynamic processes to form lattices, and the relationship between the angular frequency and the properties of number of vortices and the geometric pattern of formed lattices. In addition, similar to these previous studies, we measured the deformation parameter $\alpha$, defined below, to more precisely discuss the morphological aspects of vortex lattice formation. 
\begin{eqnarray}
\alpha \coloneqq -\Omega \frac{\braket{x^2} - \braket{y^2}}{\braket{x^2} + \braket{y^2}}.
\end{eqnarray}
where the bracket symbol $\braket{\xi}$ represents the expected value of the physical quantity $\xi$, which can be calculated as $\braket{\xi} = \int \xi |\psi|^2 dxdy$ if $\xi$ is a scalar quantity. Figure~\ref{fig:Figure5}(b) shows the time dependence of the parameter $\alpha$ for $\Omega = 0.57$, $\Omega = 0.7$, $\Omega = 0.86$, and $\Omega = 0.9$, respectively. In all cases of $\Omega$, we observed that parameter $\alpha$ gets stabilized after experiencing oscillations for a certain time. Similar behaviors were observed in the experiments~\cite{PhysRevLett.86.4443} and the simulations that utilize the higher-order finite difference method~\cite{PhysRevA.67.033610}. 
The parameter $\alpha$ was confirmed to drop abruptly to a value below 0.05 as the vortices enter the condensates when $\Omega=0.7$, similar to that reported in Ref.~\cite{PhysRevA.67.033610}. 
However, an in-depth examination revealed several differences compared to the simulation reported in Ref.~\cite{PhysRevA.67.033610}. In particular, the physical time elapsed for completing the vortex lattice formation process was approximately 735 msec for $\Omega=0.7$ in Ref.~\cite{PhysRevA.67.033610}, whereas, it was approximately 489 msec in our simulations. Thus, the vortex lattice formation process proceeded approximately 1.5 times faster in our SPH simulation. In addition, the physical time for the condensates to form an ellipsoid shape when $\Omega=0.7$ was approximately 39 msec in our case, whereas it was approximately 67 msec in Ref.~\cite{PhysRevA.67.033610}. Further, regarding the initial oscillations, it was reported that the transition from sinusoidal to aperiodic oscillations occurs at approximately $\Omega=0.75$ in Ref.~\cite{PhysRevA.67.033610}; however, in our simulation, such sinusoidal oscillations were observed only when $\Omega=0.57$. In fact, they were already non-sinusoidal in nature when $\Omega=0.7$. A possible reason for these discrepancies is that the estimation of $C_{\rm sph}$ was inaccurate because it was roughly calibrated using the measured relationship between the number of vortices and $\Omega$, as previously described. However, in our SPH scheme, $C_{\rm sph}$ is a critical factor because the magnitude of the rotational force generated by the first term in Eq.~(\ref{eq:simuleq}) is proportional to $C_{\rm sph}$. In summary, the time-dependence of $\alpha$ in our scheme is still under discussion and requires further research. 

Nevertheless, it is still significant that we could reproduce the dynamics of vortex lattice formation using the explicit time-integrating scheme and SPH, which is a Lagrangian calculation scheme. 
\HLPOFR{As mentioned in the Introduction, the GP equation is a nonlinear Schr${\rm \ddot{o}}$dinger equation for interacting bosons, and its Hamiltonian has the form of a many-particle interacting system. Now that we have succeeded in describing the GP equation in SPH, which has a form of many-particle interaction systems similar to the quantum mechanical picture, we expect to accurately model the microscopic interactions among atomic particles in future work.} \HLPOFR{In addition,} the explicit time-integration scheme is frequently preconceived as less accurate than the implicit time-integration one. Similarly, the Lagrangian calculation schemes are believed to be less accurate than the Euler approaches; for example, the finite-difference or finite-element methods. However, this study demonstrated that the Lagrangian numerical models with explicit time integration could reproduce the dynamics of vortex-lattice formation by selecting the appropriate scheme.

\HLPOFR{Furthermore}, our time-integrating scheme, which describes the GP equation in complex representation and simultaneously solves a pair of two time-dependent equations obtained for the respective parts of wavefunctions can be intuitively understood in contrast to the case of conventional methods owing to the real-time integration. As it does not require a special mathematical operation system to perform algebraic operations involving imaginary units, its implementation on a computational accelerator such as a GPU is straightforward. In fact, all calculations in this study were performed on an NVIDIA GeForce RTX2080 Ti. Thus, as our SPH scheme is a computational scheme that is compatible with high-performance computing, we can expect to explore the nonlinear dynamics of vortex lattice formation using HPC resources like GPU-rich supercomputers in the near future.

This study proposed a static SPH scheme wherein all computation points were fixed in space. This offers many advantages in view of computational physics, such as the practical ability to place calculation points anywhere in space. However, it is significant from the perspective of condensed matter physics that we reproduced vortex lattices in SPH form. A recent study suggested that a two-fluid model for superfluid helium-4 can be solved using SPH to produce vortex lattices under specific conditions; their simulations showed that a gathering of low-density components became a spinning vortex, and they form a vortex lattice eventually~\cite{doi:10.1063/5.0060605}. In contrast, this study confirmed that the branch points of the phase of wavefunctions, where density cannot be defined owning to singularities, corresponded to the holes in the vortex lattice using SPH simulations. Therefore, the results in this study and Ref.~\cite{doi:10.1063/5.0060605} are similar in that both confirmed in SPH formalism that the vortices correspond to the points where densities are vacant. Accordingly, SPH-based approaches can be helpful in determining the connectivity between these phenomena, which was observed on micro and macroscopic scales. In reality, another study theoretically presents that the motion equation for inviscid fluid in the two-fluid model becomes equal to the quantum fluid equations derived from the GP equations, in the SPH form, given that fluid exhibits moderate density change and has sufficiently slow fluid speed compared to the critical speed of superfluid helium-4~\cite{doi:10.1063/5.0122247}. Such connectivity between classical and quantum mechanics can be explored by developing a dynamic SPH model for the GP equation. By allowing the computational points to flow in our SPH scheme proposed in this study, we may expect to simultaneously solve the quantum fluid equations, that is, the macroscopic motion equations derived from the GP equations and the original GP equations in the SPH form. The higher-order scheme would be necessary to this end; nevertheless, our SPH scheme has great potential to advance both physical and computational studies on physics revolving around the dynamics of vortex lattice formation.

\section{Conclusion}
This study proposed a new numerical scheme for vortex lattice formation in a rotating BEC using smoothed particle hydrodynamics (SPH) with an explicit real-time integration scheme. Specifically, we described the Gross--Pitaevskii (GP) equation in complex representation to obtain a pair of time-dependent equations, which were solved simultaneously after discretization based on SPH particle approximation. We adopted the 4th-order Runge--Kutta method for the update; our numerical scheme has the accuracy of the 2nd order in space and the 4th order in time. 
We then performed the simulations of rotating Bose gas trapped in a harmonic potential. Consequently, our simulations are qualitatively consistent with both experiments and simulations reported in the previous studies regarding the phenomena characteristic of quantum lattices: the dynamic processes to form lattices, and the relationship between the angular frequency and the properties of the number of vortices and the geometric pattern of formed lattices. Notably, we succeeded in reproducing the geometric patterns of formed lattices for several cases; for example, the \HLPOFRR{hexagonal} lattice observed in the experiments of rotating BECs. We confirmed that the simulation began with the periodic oscillation of the condensate, which attenuated and maintained a stable rotation with slanted elliptical shapes. However, the surface was excited to be unstable and generated ripples, which grew into vortices and then penetrated the inside the condensate, eventually forming a lattice. We confirmed that each branch point of the phase of wavefunctions corresponds to each vortex. 
The same observations were reported in experiments and the simulations using the higher-order finite difference scheme, demonstrating that our approach exhibited a certain degree of accuracy. In conclusion, we reproduced the dynamics of vortex lattice formation using SPH with the explicit time-integrating scheme and real-time update; we successfully developed a new Lagrangian method for the simulations of vortex lattice formation in rotating BECs. 







\section*{Acknowledgment}
This study was supported by JSPS KAKENHI Grant Number 22K14177. 
The author would like to thank Editage (www.editage.jp) for English language editing.
The author specially acknowledges Prof. Katsuhiro Nishinari and the administrative staff at the Nishinari Laboratory and the RCAST, University of Tokyo.
The author is also grateful to his family for their warm encouragement.

\bibliographystyle{h-physrev3}
\bibliography{reference}

\begin{thebibliography}{10}

\bibitem{PhysRevE.62.1382}
M.~M. Cerimele, M.~L. Chiofalo, F.~Pistella, S.~Succi, and M.~P. Tosi,
\newblock Numerical solution of the gross-pitaevskii equation using an explicit
  finite-difference scheme: An application to trapped bose-einstein
  condensates, Phys. Rev. E {\bf 62}, 1382 (2000).

\bibitem{PhysRevE.62.2937}
S.~K. Adhikari,
\newblock Numerical study of the spherically symmetric gross-pitaevskii
  equation in two space dimensions, Phys. Rev. E {\bf 62}, 2937 (2000).

\bibitem{PhysRevE.62.7438}
M.~L. Chiofalo, S.~Succi, and M.~P. Tosi,
\newblock Ground state of trapped interacting bose-einstein condensates by an
  explicit imaginary-time algorithm, Phys. Rev. E {\bf 62}, 7438 (2000).

\bibitem{VERGEZ2016144}
G.~Vergez, I.~Danaila, S.~Auliac, and F.~Hecht,
\newblock A finite-element toolbox for the stationary
  gross^^e2^^80^^93pitaevskii equation with rotation, Computer Physics
  Communications {\bf 209}, 144 (2016).

\bibitem{doi:10.1137/15M1009172}
P.~Henning and A.~M\r{a}lqvist,
\newblock The finite element method for the time-dependent gross--pitaevskii
  equation with angular momentum rotation, SIAM Journal on Numerical Analysis
  {\bf 55}, 923 (2017), https://doi.org/10.1137/15M1009172.

\bibitem{HEID2021110165}
P.~Heid, B.~Stamm, and T.~P. Wihler,
\newblock Gradient flow finite element discretizations with energy-based
  adaptivity for the gross-pitaevskii equation, Journal of Computational
  Physics {\bf 436}, 110165 (2021).

\bibitem{doi:10.1063/1.4887568}
Y.~Y. Choy, W.~N. Tan, K.~G. Tay, and C.~T. Ong,
\newblock Crank-nicolson implicit method for the nonlinear schrodinger equation
  with variable coefficient, AIP Conference Proceedings {\bf 1605}, 76 (2014),
  https://aip.scitation.org/doi/pdf/10.1063/1.4887568.

\bibitem{MURUGANANDAM20091888}
P.~Muruganandam and S.~Adhikari,
\newblock Fortran programs for the time-dependent gross--pitaevskii equation in
  a fully anisotropic trap, Computer Physics Communications {\bf 180}, 1888
  (2009).

\bibitem{WANG20161114}
P.~Wang and C.~Huang,
\newblock Split-step alternating direction implicit difference scheme for the
  fractional schr^^c3^^b6dinger equation in two dimensions, Computers \&
  Mathematics with Applications {\bf 71}, 1114 (2016).

\bibitem{LI201538}
L.~Z. Li, H.-W. Sun, and S.-C. Tam,
\newblock A spatial sixth-order alternating direction implicit method for
  two-dimensional cubic nonlinear schr^^c3^^b6dinger equations, Computer
  Physics Communications {\bf 187}, 38 (2015).

\bibitem{LEHTOVAARA2007148}
L.~Lehtovaara, J.~Toivanen, and J.~Eloranta,
\newblock Solution of time-independent schr^^c3^^b6dinger equation by the
  imaginary time propagation method, Journal of Computational Physics {\bf
  221}, 148 (2007).

\bibitem{doi:10.1063/1.4821126}
P.~Bader, S.~Blanes, and F.~Casas,
\newblock Solving the schr^^c3^^b6dinger eigenvalue problem by the imaginary
  time propagation technique using splitting methods with complex coefficients,
  The Journal of Chemical Physics {\bf 139}, 124117 (2013),
  https://doi.org/10.1063/1.4821126.

\bibitem{GOLDBERG1967433}
A.~Goldberg and J.~L. Schwartz,
\newblock Integration of the schr^^c3^^b6dinger equation in imaginary time,
  Journal of Computational Physics {\bf 1}, 433 (1967).

\bibitem{PhysRevLett.82.4956}
D.~L. Feder, C.~W. Clark, and B.~I. Schneider,
\newblock Vortex stability of interacting bose-einstein condensates confined in
  anisotropic harmonic traps, Phys. Rev. Lett. {\bf 82}, 4956 (1999).

\bibitem{PhysRevA.61.011601}
D.~L. Feder, C.~W. Clark, and B.~I. Schneider,
\newblock Nucleation of vortex arrays in rotating anisotropic bose-einstein
  condensates, Phys. Rev. A {\bf 61}, 011601 (1999).

\bibitem{Muruganandam_2003}
P.~Muruganandam and S.~K. Adhikari,
\newblock Bose^^e2^^80^^93einstein condensation dynamics in three dimensions by
  the pseudospectral and finite-difference methods, Journal of Physics B:
  Atomic, Molecular and Optical Physics {\bf 36}, 2501 (2003).

\bibitem{Rogel-Salazar_2013}
J.~Rogel-Salazar,
\newblock The gross--pitaevskii equation and bose--einstein condensates,
  European Journal of Physics {\bf 34}, 247 (2013).

\bibitem{Salasnich2017BSIUA}
L.~Salasnich,
\newblock Bright solitons in ultracold atoms, Optical and Quantum Electronics
  {\bf 49}, 409 (2017).

\bibitem{doi:10.1063/5.0122247}
S.~Tsuzuki,
\newblock Theoretical framework bridging classical and quantum mechanics for
  the dynamics of cryogenic liquid helium-4 using smoothed-particle
  hydrodynamics, Physics of Fluids {\bf 34}, 127116 (2022),
  https://doi.org/10.1063/5.0122247.

\bibitem{Idowu2001}
O.~C. Idowu, D.~Kivotides, C.~F. Barenghi, and D.~C. Samuels,
\newblock {\em Numerical Methods for Coupled Normal-Fluid and Superfluid Flows
  in Helium II} (Springer Berlin Heidelberg, Berlin, Heidelberg, 2001), pp.
  162--176.

\bibitem{doi:10.1063/1.4828892}
A.~W. Baggaley and S.~Laizet,
\newblock Vortex line density in counterflowing he ii with laminar and
  turbulent normal fluid velocity profiles, Physics of Fluids {\bf 25}, 115101
  (2013), https://doi.org/10.1063/1.4828892.

\bibitem{PhysRevLett.120.155301}
S.~Yui, M.~Tsubota, and H.~Kobayashi,
\newblock Three-dimensional coupled dynamics of the two-fluid model in
  superfluid $^{4}\mathrm{He}$: Deformed velocity profile of normal fluid in
  thermal counterflow, Phys. Rev. Lett. {\bf 120}, 155301 (2018).

\bibitem{doi:10.1063/1.5091567}
C.~L. Horner and R.~A. Van~Gorder,
\newblock Dynamics of quantized vortex filaments under a local induction
  approximation with second-order correction, Physics of Fluids {\bf 31},
  065103 (2019), https://doi.org/10.1063/1.5091567.

\bibitem{PhysRevLett.124.155301}
S.~Yui, H.~Kobayashi, M.~Tsubota, and W.~Guo,
\newblock Fully coupled two-fluid dynamics in superfluid $^{4}\mathrm{He}$:
  Anomalous anisotropic velocity fluctuations in counterflow, Phys. Rev. Lett.
  {\bf 124}, 155301 (2020).

\bibitem{gingold1977smoothed}
R.~A. Gingold and J.~J. Monaghan,
\newblock Smoothed particle hydrodynamics: theory and application to
  non-spherical stars, Monthly notices of the royal astronomical society {\bf
  181}, 375 (1977).

\bibitem{Liu2010}
M.~B. Liu and G.~R. Liu,
\newblock Smoothed particle hydrodynamics (sph): an overview and^^c2^^a0recent
  developments, Archives of Computational Methods in Engineering {\bf 17}, 25
  (2010).

\bibitem{imoto2019convergence}
Y.~Imoto, S.~Tsuzuki, and D.~Nishiura,
\newblock Convergence study and optimal weight functions of an explicit
  particle method for the incompressible navier--stokes equations,
  Computational Particle Mechanics {\bf 6}, 671 (2019).

\bibitem{doi:10.1063/1.5068697}
T.~Ye, D.~Pan, C.~Huang, and M.~Liu,
\newblock Smoothed particle hydrodynamics (sph) for complex fluid flows: Recent
  developments in methodology and applications, Physics of Fluids {\bf 31},
  011301 (2019).

\bibitem{doi:10.1063/5.0060605}
S.~Tsuzuki,
\newblock Reproduction of vortex lattices in the simulations of rotating liquid
  helium-4 by numerically solving the two-fluid model using smoothed-particle
  hydrodynamics incorporating vortex dynamics, Physics of Fluids {\bf 33},
  087117 (2021), https://doi.org/10.1063/5.0060605.

\bibitem{PhysRevA.67.033610}
K.~Kasamatsu, M.~Tsubota, and M.~Ueda,
\newblock Nonlinear dynamics of vortex lattice formation in a rotating
  bose-einstein condensate, Phys. Rev. A {\bf 67}, 033610 (2003).

\bibitem{PhysRevA.65.023603}
M.~Tsubota, K.~Kasamatsu, and M.~Ueda,
\newblock Vortex lattice formation in a rotating bose-einstein condensate,
  Phys. Rev. A {\bf 65}, 023603 (2002).

\bibitem{Berman2018}
P.~R. Berman,
\newblock {\em Mathematical Preliminaries} (Springer International Publishing,
  Cham, 2018), pp. 33--51.

\bibitem{monaghan1992smoothed}
J.~J. Monaghan,
\newblock Smoothed particle hydrodynamics, Annual review of astronomy and
  astrophysics {\bf 30}, 543 (1992).

\bibitem{koshizuka1996moving}
S.~Koshizuka and Y.~Oka,
\newblock Moving-particle semi-implicit method for fragmentation of
  incompressible fluid, Nuclear science and engineering {\bf 123}, 421 (1996).

\bibitem{Koshizuka1998}
S.~Koshizuka, A.~Nobe, and Y.~Oka,
\newblock Numerical analysis of breaking waves using the moving particle
  semi-implicit method, International Journal for Numerical Methods in Fluids
  {\bf 26}, 751 (1998).

\bibitem{Tsuzuki_2021}
S.~Tsuzuki,
\newblock Particle approximation of the two-fluid model for superfluid 4he
  using smoothed particle hydrodynamics, Journal of Physics Communications {\bf
  5}, 035001 (2021).

\bibitem{PhysRevLett.86.4443}
K.~W. Madison, F.~Chevy, V.~Bretin, and J.~Dalibard,
\newblock Stationary states of a rotating bose-einstein condensate: Routes to
  vortex nucleation, Phys. Rev. Lett. {\bf 86}, 4443 (2001).

\bibitem{4541126}
D.~Luebke,
\newblock Cuda: Scalable parallel programming for high-performance scientific
  computing,
\newblock in {\em 2008 5th IEEE International Symposium on Biomedical Imaging:
  From Nano to Macro}, pp. 836--838, 2008.

\bibitem{GREST1989269}
G.~S. Grest, B.~D^^c3^^bcnweg, and K.~Kremer,
\newblock Vectorized link cell fortran code for molecular dynamics simulations
  for a large number of particles, Computer Physics Communications {\bf 55},
  269 (1989).

\bibitem{GOMEZGESTEIRA2012289}
M.~Gomez-Gesteira {\em et~al.},
\newblock Sphysics ^^e2^^80^^93 development of a free-surface fluid solver
  ^^e2^^80^^93 part 1: Theory and formulations, Computers \& Geosciences {\bf
  48}, 289 (2012).

\bibitem{ramachandran2013pysph}
P.~Ramachandran and K.~Puri,
\newblock Pysph: {A} framework for parallel particle simulations, {In}
  proceedings of the 3rd {International} {Conference} on {Particle}-{Based}
  {Methods} ({Particles} 2013), {Stuttgart}, {Germany},
\newblock 18th, 2013.

\bibitem{KISHORKUMAR201974}
R.~{Kishor Kumar}, V.~Lon\u{c}ar, P.~Muruganandam, S.~K. Adhikari, and
  A.~Bala\u{z},
\newblock C and fortran openmp programs for rotating bose--einstein
  condensates, Computer Physics Communications {\bf 240}, 74 (2019).

\end{thebibliography}

\end{document}